\documentclass{PoS}

\usepackage{amsmath}
\usepackage{amsfonts}
\usepackage{amssymb}
\usepackage{graphicx} 
\usepackage{cite}

\newcommand{\eq}[1]{\begin{equation}
                     \begin{split} #1 \end{split}
                     \end{equation}}
\newcommand{\ov}{\overline}
\newcommand{\op}{\hspace{1pt}}

\title{Large field inflation and string moduli stabilization}

\ShortTitle{Large field inflation and string moduli stabilization}

\author{\speaker{Ralph Blumenhagen}, Michael Fuchs, Daniela Herschmann\\
        Max-Planck-Institut f\"ur Physik (Werner-Heisenberg-Institut), \\ 
   F\"ohringer Ring 6,  80805 M\"unchen, Germany \\
        E-mail: \email{blumenha,mfuchs,herschma@mpp.mpg.de}}

\author{Anamaria Font\\
Departamento de F\'{\i}sica, Centro de F\'{\i}sica Te\'orica y Computacional \\
Facultad de Ciencias, Universidad Central de Venezuela\\
 A.P. 20513, Caracas 1020-A, Venezuela\\
        E-mail: \email{afont@fisica.ciens.ucv.ve}}

\author{Erik Plauschinn\\
        Arnold-Sommerfeld-Center f\"ur Theoretische Physik \\
        Ludwig-Maximilians-Universit\"at M\"unchen \\ 
        Theresienstra\ss e 37, 80333 M\"unchen, Germany\\
        E-mail: \email{erik.plauschinn@lmu.de}}

\abstract{Recent developments on large scale string moduli stabilization via non-geometric
fluxes are reviewed. In the framework of type IIB  orientifolds on fluxed Calabi-Yau 
manifolds, the existence of so-called flux-scaling minima provides
a good starting point for the realization of parametrically controlled
models of F-term axion monodromy inflation.}

\FullConference{18th International Conference From the Planck Scale to the Electroweak Scale \\
		25-29 May 2015\\
		Ioannina, Greece }

\begin{document}

\section{Introduction}

The motivation of the work reviewed in this article  lies in two recent experimental inputs. 
Firstly, the initial run of the LHC did not give evidence for supersymmetry. 
However, most approaches in string phenomenology have focused on designing models
with a low supersymmetry breaking scale of around 1 TeV in the observable sector.
In fact, TeV scale supersymmetry breaking was used to  fix the other mass
scales relevant in string theory compactification, such as 
the string scale itself or the Kaluza-Klein scale. 
Thus, having a large hierarchy relative to the Planck scale of around 15 orders
of magnitude, exponential hierarchies seemed to be a very natural starting point. 
This was achieved, for instance,  in the KKLT \cite{Kachru:2003aw } or in the large volume scenario \cite{Balasubramanian:2005zx,Conlon:2005ki} by utilizing non-perturbative contributions to the superpotential 
coming from instantons or gaugino condensates. 
Since string theory generically only predicts supersymmetry to be present
at the string scale, in general there is no model independent prediction
for the supersymmetry breaking scale in the observable standard model (SM) like sector. Thus,
one can also contemplate a scheme of moduli stabilization, in which 
the susy breaking scale comes out much higher than the TeV scale.

Secondly, in March 2014 the BICEP2 collaboration 
announced the detection of primordial B-modes with a
remarkably large tensor-to-scalar ratio of $r \sim 0.2$. Although 
meanwhile this effect was clarified to be mainly induced by  foreground dust, it motivated much research in string cosmology. 
So far, most string inspired models of inflation predicted  lower tensor-to-scalar ratios.
For a ratio of $r> 0.01$  the Lyth bound \cite{Lyth:1996im} implies 
that the inflaton has to run  over trans-Planckian field ranges, hence  making the process highly UV sensitive. 
Therefore, string theory as a UV complete theory of gravity
provides a well defined framework to discuss high scale inflation.

In particular, to forbid higher order Planck suppressed operators in the inflaton action,
one can employ  a pseudo-scalar field with a continuous shift symmetry, called an axion. 
This approach is especially appealing for string theory in which axions appear naturally from compactifying $p$-forms to four dimensions. 
There are two important classes of models for axionic inflation in string theory. 
The first uses the periodic cosine potential \cite{Freese:1990rb} generically generated by instantons, possibly with more than one axion to enlarge the field range \cite{Kim:2004rp,Dimopoulos:2005ac}. 
There is an ongoing debate in the literature whether it is possible to have a trans-Planckian field range in these models. 
They face some challenges in being consistent with the weak gravity conjecture \cite{ArkaniHamed:2006dz}. 
In our work we focus on a  second approach that  uses a controlled weak breaking of the axionic shift symmetry \cite{Kaloper:2008fb}. 
The cosine potential from the instanton is still present but over every period the potential energy increases by 
a certain amount.  The cosine potential is therefore only  a modulation on top of the potential that drives inflation. This ansatz is called axion monodromy inflation and was introduced in the stringy context \cite{Silverstein:2008sg}.

One mechanism  to generate  a polynomial potential for axion monodromy inflation 
is to turn on background fluxes generating a tree-level F-term scalar potential \cite{Marchesano:2014mla,Blumenhagen:2014gta,Hebecker:2014eua},
 see also \cite{Palti:2014kza,Grimm:2014vva,Ibanez:2014kia,Arends:2014qca,Hassler:2014mla,McAllister:2014mpa,Ibanez:2014swa,Buchmuller:2015oma,Retolaza:2015sta,Bielleman:2015ina} and for reviews \cite{Baumann:2014nda,Westphal:2015eva}. 
For other attempts to realize axion monodromy inflation in string theory see e.g. \cite{Hebecker:2014kva,Escobar:2015fda}.
Turning on fluxes has the advantage that the same mechanism  generating the axion potential also stabilizes the other moduli and breaks supersymmetry. One starts with e.g. an orientifolded Calabi-Yau compactification of the 10-dimensional type IIA or type IIB theory giving rise to a four dimensional $N = 1$ supergravity theory with usually plenty of massless scalar fields and axions, altogether called moduli. This geometry is then perturbed by turning on vacuum expectation values of the field strengths on the Calabi-Yau. These ``fluxes'' induce a polynomial potential for scalar fields giving them a mass. The BICEP2 results  and the so far not detected non-gaussianities would have favored a model of chaotic inflation with a single  scalar field 
governed by a purely quadratic potential. 
In this case, the potential energy  during inflation is $M_{\rm inf} \sim 10^{16}$ GeV, the Hubble-scale during inflation is $H_{\rm inf} \sim 10^{14}$ GeV and the inflaton mass is $M_{\theta} \sim 10^{13}$ GeV. In order to use an effective supergravity approach, 
the string scale $M_{\rm s}$ and the Kaluza-Klein scale $M_{\rm KK}$ must lie
above all these scales.
Moreover,  the other moduli masses should lie above the Hubble scale to guarantee a model of single field inflation. Therefore, altogether we have the ordered hierarchy of mass scales
\eq{
\label{introhierarchy}
M_{\rm Pl} \; > \; M_{\rm s} \; > \; M_{\rm KK} \; > \; M_{\rm inf} \; \sim \;M_{\rm mod} \; > \; H_{\rm inf} \; > \; M_{\rm \theta } \,,
}
where neighboring scales can differ only by a factor of ${\mathcal O}(10)$. This is obviously a major challenge for concrete string model building and it requires a scheme of moduli
stabilization with mass scales closer to the GUT scale than to the TeV scale.
Of course this could  happen just  by a numerical coincidence, but parametrically controlling this hierarchy would be more desirable. 
The first challenge is to get only one axion lighter than the other moduli. This is difficult, as usually the axions are paired together with another  scalar, the  so-called saxion, into a
complex scalar. 
After moduli stabilization, i.e. at the minimum of the scalar potential,
usually the masses turn out to be of the same order and especially to separate the mass of the inflaton from its saxionic partner is expected to be difficult to achieve.

To address this question we follow a procedure developed by some of the authors in \cite{Blumenhagen:2014gta, Blumenhagen:2014nba}. In a first step one uses large fluxes to stabilize all moduli except a single  axion, which will be the inflaton candidate. In the second step one stays in the minimum and turns on additional fluxes of ${\mathcal O}(1)$ to stabilize the 
remaining axion which is then parametrically lighter than the rest. Potential issues are that the existing minimum can be destabilized by the additional fluxes, and that the large fluxes backreact onto the geometry. In the original papers \cite{Blumenhagen:2014gta ,Blumenhagen:2014nba}, 
examples were constructed that indeed satisfy these criteria. However, in these models the K\"ahler moduli were 
initially neglected, and only 
stabilized in a second step by subleading effects coming from exponentially suppressed instanton
corrections. The masses of these K\"ahler moduli are therefore generically lighter 
than the tree-level induced inflaton mass. A problem in these constructions is that the fluxes cannot be tuned 
arbitrarily small due to quantization conditions. However, this constraint can be evaded as we will explain.

To recapitulate, both the possibility of having a higher scale of supersymmetry breaking and
a large tensor-to-scalar ratio with high scale
inflation, motivated us  in \cite{Blumenhagen:2015kja} to consider a parametrically
controlled  scheme of moduli stabilization where only polynomial hierarchies appear. 
For this purpose, all moduli including the K\"ahler moduli should be stabilized at a high scale by a tree-level flux induced potential. For the complex structure moduli and the axio-dilaton it is well known how to achieve this in type IIB string theory with the usual geometric fluxes, 
namely the $H = dB$ flux corresponding to the Kalb-Ramond $B$ field, the Ramond-Ramond three-form flux ${\mathfrak F}^{(3)}$, and the geometric flux $F$ which encodes deformation of the background  metric. 
However,  to stabilize the K\"ahler moduli at tree-level we need to turn on also 
non-geometric $Q$- and $R$-fluxes. The latter can arise when applying T-duality transformations
to a background with the geometric $H$ and $F$ fluxes \cite{Shelton:2005cf}. 

\begin{figure}[t] \label{Tfold}
\centering
\includegraphics[scale=0.5]{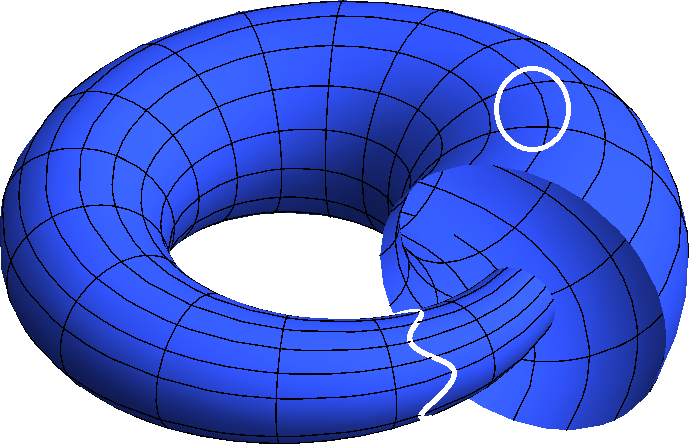}
\caption{A picture of a non-geometric T-fold where T-dualities with their characteristic inversion of the radius can be used as transition functions. A global description as a manifold fails on a T-fold. When a string passes the step its momentum becomes winding. Thus a propagating string becomes a wounded string as depicted. }
\end{figure}

T-duality is a purely stringy symmetry which inverts the volume in a chosen isometry direction and  interchanges the momentum with the winding number of the string. 
For non-geometric backgrounds not only geometric transition functions such as diffeomorphisms and gauge transformations are allowed, but also T-duality- and $\beta$-transformations. These transformations mix the metric and the $B$-field. Due to the inversion of the volume and the mixing of metric and $B$-field when going from one patch to another, a smooth geometric description of a manifold fails. On the T-fold where only the $Q$-flux is turned on, a local description with a metric is still possible and only the global transition functions are non-geometric \cite{Hull:2004in}. An illustration of the prototype example is shown in figure \ref{Tfold}. 
The nature of the $R$-flux backgrounds is less clear, as it arises from a usually forbidden T-duality in a non-isometry direction. The $R$-flux was first argued for using duality arguments between the T-dual IIA and IIB theory and becomes necessary for writing down  a
duality invariant superpotential \cite{Shelton:2005cf}.
Is it expected that here even a local geometric description fails
and that the space becomes non-associative \cite{Bouwknegt:2004ap,Blumenhagen:2010hj,Lust:2010iy}.  

Since we are going to employ non-geometric fluxes for our models, we would like to mention the 
following known issues:
\begin{itemize}
\item For moduli stabilization we work in the effective four-dimensional supergravity theory. 
From the four-dimensional point of view, our solutions are completely unambiguous. 
Usually, when  finding solutions to the four-dimensional theory  one hopes that this points towards genuine solutions of the full ten-dimensional theory. As shown by some of the authors in \cite{Blumenhagen:2015lta}, the ten-dimensional theory behind our setup is not the standard IIB supergravity, but double field theory (DFT) \cite{Siegel:1993th,Hull:2009mi,Hohm:2010jy,Hohm:2010pp}. 
More concretely, the scalar potential we are using results from an orientifold
projection of DFT compactified on a Calabi-Yau manifold
equipped with constant (non-)geometric fluxes. This scalar potential
is nothing else than the orientifolded potential of ${\cal N}=2$ gauged supergravity (GSUGRA).
Although derived from string theory, the uplift of solutions of double field theory to string theory is only guaranteed for solutions obeying a section condition. But double field theory with this section condition reduces to the NS-NS part of the superstring which does not contain non-geometric fluxes. 
There are attempts to relax this section condition \cite{Blumenhagen:2014gva, Blumenhagen:2015zma} but we cannot assert that every solution constructed here uplifts to string theory.

\item In non-geometric backgrounds large and small volumes are identified. This makes it difficult to define a 
Kaluza-Klein scale, which explicitly depends on the volume, and to determine whether this scale separates 
from the moduli scale especially as the difference should be only of 
${\mathcal O}(1)$. We will nonetheless calculate the KK scale with the traditional formula. We take the point of view that the fluxes we are turning on are only a small correction to the existing geometry. 
\end{itemize}

Keeping in mind these questions, we studied the vacuum structure  of the scalar potential. In order to work with a small number of moduli, we were not careful about the specification of a concrete CY geometry and orientifold projection, but only assumed  the existence of manifold with the required topological data.
Having parametrically controlled axion F-term monodromy inflation as the ultimate goal, we headed  for a subclass of models where the following aspects are realized:
\begin{itemize}
\item The vacua should be tachyon free, so that after uplifting they can lead to 
metastable de Sitter vacua.
\item There exists a no-go theorem \cite{Conlon:2006tq} which states that in a 
supersymmetric vacuum, when an axion is completely unstabilized, its saxionic partner is either massless or tachyonic. Since we usually leave one axion initially unstabilized for an eventual inflaton candidate, we searched  for non-supersymmetric vacua. 
\item The moduli should be stabilized in their perturbative regime, thus at large radius, large complex structure and weak string coupling.
\item All saxionic moduli should be stabilized with axions providing candidates for the inflaton. 
\item The moduli masses are smaller than the string and the Kaluza-Klein scale. 
\end{itemize}

This proceedings article is organized as follows. In section 2 we  describe the setup of our models. We comment on the ten-dimensional double field theory origin and on the reduced theory in four dimensions. In section 3 we  present  a concrete toy model which
 exemplifies  the features of our class of models. We  also comment on phenomenological aspects such as the hierarchies, tachyon uplift, sequestering and de Sitter uplift. Section 4 is dedicated to inflation. We first discuss
how to gain parametric control over an axions mass by using the two step procedure described above. 
Then, we take the toy model of section 3 and investigate inflation in just one step.
Section 5 contains a summary and conclusions.
This review article is based on the recent publications 
\cite{Blumenhagen:2014gta, Blumenhagen:2015lta,Blumenhagen:2015qda,Blumenhagen:2015kja,Blumenhagen:2015xpa}.


\section{The setup of orientifolded ${\cal N}=2$ GSUGRA}

To obtain an effective four-dimensional string theory it is necessary to compactify six dimensions. 
These six internal dimensions are specified by their geometry, and by the expectation values of
the various field strengths of the theory -- the so-called fluxes. These parameters determine the physics in the effective four dimensional field theory. In  type IIB string theory, the closed-string fluxes are given by 
the Neveu-Schwarz--Neveu-Schwarz (NS-NS) field strength $H=dB$, 
the so-called geometric flux $F$, and the Ramond-Ramond (R-R) flux $\mathfrak F^{(3)}$. However, 
it has been  argued  via T-duality that also non-geometric fluxes $Q$ and $R$ can be present \cite{Shelton:2005cf}.
In ordinary string theory it is not known how to consistently describe geometric and
non-geometric fluxes in a unified way; double field theory (DFT) on the other hand provides 
such a framework.

The main idea of double field theory is to double  the number of space-time coordinates
by extending the $D$ usual coordinates by $D$ so-called winding coordinates
\cite{Siegel:1993th,Hull:2009mi,Hohm:2010jy,Hohm:2010pp}.  
On such a space it is possible to formulate a generalized
gravity theory that is manifestly $O(D,D)$ invariant and covariant with respect
to generalized diffeomorphisms.  
In particular, T-duality is now  a manifest symmetry as it is  part of the $O(D,D)$ symmetry group.  For closure of the symmetries and consistency of the theory, 
one is forced to require a so-called section- or strong-constraint that effectively
projects  the doubled $2D$  dimensional space  onto a $D$ dimensional slice.   
Eliminating in this manner the winding coordinates one gets back the supergravity
theory.

The connection between DFT and
gauged supergravity has first been made in \cite{Aldazabal:2011nj, Geissbuhler:2011mx}, where it has been shown that compactification of
DFT on a fluxed six-torus leads to ${\cal N}=4$ gauged supergravity in four dimensions. 
In the framework of generalized geometry also fluxed CY manifolds 
with  $SU(3) \times SU(3)$ structure 
were considered \cite{Grana:2005ny, Benmachiche:2006df, Grana:2006hr}.
In the recent work \cite{Blumenhagen:2015lta} DFT was compactified on  fluxed
Calabi-Yau (CY) three-folds, which result in  gauged  ${\cal N}=2$ supergravity 
theories  in four dimensions. 
There it was shown that the DFT Lagrangian in the flux formulation
can be expressed on a fluxed CY in the following way
\eq{ \label{onlyyou}
  \star \mathcal L_{\rm NS\op NS} = -e^{-2\phi}\bigg[\hspace{18pt}
  &
  {1\over 2}\, \chi \wedge \star\op \ov\chi \;+\;{1\over 2}\op \Psi\wedge \star\op \ov\Psi \\
  -& {1\over 4}\op \big(\Omega_3\wedge \chi\big)\wedge\star \op \big(\ov\Omega_3\wedge \ov \chi\big)
  -{1\over 4}\op \big(\Omega_3\wedge \ov\chi\big)\wedge\star\op \big(\ov\Omega_3\wedge  \chi\big)
  \hspace{8pt}\biggr]\,,
}
where $\chi={\mathfrak D} \op  e^{iJ}$ and $\Psi=\mathfrak D\op
\Omega_3$. 
Here $J$ and $\Omega_3$ are the K\"ahler  and holomorphic three-form 
of the Calabi-Yau three-fold, and ${\mathfrak D}$ denotes a twisted differential 
that involves all  geometric and non-geometric fluxes. The latter is defined as
\eq{
   \mathfrak D = d - \mathfrak H\wedge\: - 
  \mathfrak F\circ\: - \mathfrak Q\bullet\: - \mathfrak R\,\llcorner \,,
}
in which the flux orbits \cite{Blumenhagen:2013hva} are given by (see also
\cite{Gao:2015nra})
\eq{
\label{orbitfluxes}
\mathfrak{H}_{ijk}&={H}_{ijk}+3\op {F}^m{}_{[\underline{ij}}\, B_{m\underline{k}]}
    +3\, {Q}_{[\underline{i}}{}^{mn} B_{m\underline{j}}\op  B_{n\underline{k}]}
    +{R}^{mnp}  B_{m[\underline{i}} \op B_{n\underline{j}}\op  B_{p\underline{k}]} \,, 
\\
\mathfrak{F}^i{}_{jk}&={F}^i{}_{jk}+2\,{Q}_{[\underline{j}}{}^{mi}\op  B_{m\underline{k}]} 
    +{R}^{mni}  B_{m[\underline{j}} \op B_{n\underline{k}]} \,,
\\
\mathfrak{Q}_k{}^{ij}&={Q}_k{}^{ij}+{R}^{mij}  B_{mk} \,, 
\\
\mathfrak{R}^{ijk}&={R}^{ijk}\,,
}
where $B_{ij}$ are the components of the Kalb-Ramond field along the internal manifold.
These tensor fields can be regarded as operators acting on differential forms. In particular,
employing a local basis $\{dx^i\}$ and the contraction $\iota_i$ satisfying $\iota_i\op dx^j = \delta_i^j$, 
this action can be implemented by 
\eq{
  \label{flux_ops}
  \arraycolsep1pt
  \begin{array}{l@{\hspace{4pt}}c@{\hspace{3pt}}cc@{\hspace{5pt}}cccccl}
  H \,\wedge &=&  \frac{1}{3!} & H_{ijk} & dx^i &\wedge& dx^j  &\wedge& dx^k & \,, \\[6pt]
  F \,\circ &=&  \frac{1}{2!} & F^k{}_{ij} &  dx^i  &\wedge& dx^j & \wedge & \iota_k & \,,   \\[6pt]
  Q \,\bullet &=&  \frac{1}{2!} & Q_i{}^{jk} & dx^i &\wedge& \iota_j  &\wedge& \iota_k & \,,   \\[6pt]  
  R \,\llcorner &=& \frac{1}{3!} & R^{ijk} & \iota_i &\wedge& \iota_j &\wedge& \iota_k & \,.   
  \end{array}
}
Since the Lagrangian \eqref{onlyyou} does not contain the unknown metric of the CY
explicitly, but only in combinations that can be expressed in terms of the
complex structure and K\"ahler moduli, integration over the internal coordinates
was straightforward. This led precisely to the form of the scalar potential of 
${\cal N}=2$ gauged supergravity, as it was expressed in \cite{D'Auria:2007ay}. 

As mentioned above, compactifications on Calabi-Yau three-folds preserve ${\cal N}=2$ supersymmetry in  four dimensions. In order to obtain an ${\cal N}=1$ theory, one performs an orientifold projection under which
part of the fields in the theory are projected out. For definiteness, here we choose the 
projection such that it leads to orientifold three- and seven-planes. The remaining fields
in the K\"ahler moduli sector can be expressed as
\eq{
  \label{moduli}
  \begin{array}{l@{\hspace{3.5pt}}l}
  S &=  e^{-\phi} -i\op C^{(0)} := s + i\op c  \,, \\[10pt]
  G^a &= S\, b^a + i\op c^a \,, \\[6pt] 
  T_\alpha & \displaystyle =\tau_{\alpha}
   +i\op \rho_\alpha-\frac{i}{2}\op\kappa_{\alpha a b} \op c^a b^b
  -\frac{1}{4} \op e^\phi \op \kappa_{\alpha a b} \op {G}^a (G +\ov G)^b 
  \,,
  \end{array}
}
where $\phi$ is the dilaton, $C^{(0)}$ denotes the R-R zero-form potential, and
$\kappa_{\alpha\beta\gamma}$ and $\kappa_{\alpha ab}$ are triple intersection numbers of the 
Calabi-Yau manifold $\mathcal X$. Furthermore, $t^{\alpha}$, $c^a$ and $b^a$ are the components of the K\"ahler form, $C^{(2)}$ and $B$ along the internal space; $\rho_{\alpha}$ are the corresponding components of $C^{(4)}$ and $\tau_{\alpha}=\frac{1}{2}\,\kappa_{\alpha\beta\gamma}\op t^\beta  t^\gamma$.
The indices label elements in the orientifold-even and -odd cohomologies as
$\alpha = 1,\ldots, h^{1,1}_+$ and $a = 1, \ldots, h^{1,1}_-$, and in the following we denote 
two-forms and four-forms on the Calabi-Yau three-fold $\mathcal X$ by $\omega_{\star}$ and $\tilde\omega^{\star}$.
For later reference, we also arrange the moduli fields into an even complex multi-form as
\eq{
  \label{def_01}
  \Phi^{\rm ev}_c = i\op S -i\op G^a\op \omega_a -i\op T_{\alpha}\op \tilde\omega^{\alpha} \,.
}
Turning to the complex-structure moduli sector, we mention that the holomorphic three-form 
$\Omega_3\in H^3_-(\mathcal X)$ can 
be expanded as 
\eq{
  \label{exp_02}
  \Omega_3 = X^{\lambda} \alpha_{\lambda} - F_{\lambda} \op\beta^{\lambda} \,,
}
where $\lambda = 0, \ldots, h^{2,1}_-$ and $\{\alpha_{\lambda},\beta^{\lambda}\}$ denotes a symplectic basis. 
Note that the periods $F_{\lambda}=  \partial F/\partial X^{\lambda}$ depend on 
$X^{\lambda}$ via a holomorphic prepotential that in the large complex structure
regime takes the simple form $F = \frac{d_{ijk} X^{i}X^{j}X^{k}}{X^0}$.
In general the prepotential   is subject to perturbative and non-perturbative corrections,
which take the general form (see for instance
\cite{Hosono:1994av})
\eq{
  \label{f_corr}
  \widetilde F = F + \frac{1}{2} \op a_{ij} X^i X^j + b_i X^i X^0 + \frac{1}{2}\op i\op\gamma \bigl( X^0\bigr)^2
  + F_{\rm inst.} \,,
}
where the constants $a_{ij}$ and $b_i$ are rational  numbers, 
while $\gamma$ is real. 
Moreover, the complex-structure moduli are given by
\eq{
  U^i =v^i + i\op u^i = - i \op \frac{X^i}{X^0} \,, \hspace{60pt} i = 1, \ldots, h^{2,1}_- \,.
}
The four-dimensional ${\cal N}=1$ theory after the orientifold projection is described by the following K\"ahler potential \cite{Grimm:2004uq}
\eq{
  \label{kpot}
  K = -\log\bigl[ \,S+\ov S\, \bigr] - 2\log {\mathcal V}
  -\log \left[ \; i \int_{\mathcal X} \Omega\wedge \ov\Omega \:
  \right] \,,
}
where ${\mathcal V} = \frac{1}{3!} \,\kappa_{\alpha\beta\gamma}\op  t^{\alpha} t^{\beta} t^{\gamma}$ 
denotes the volume of the Calabi-Yau three-fold in Einstein frame.
The superpotential in the presence of 
R-R three-form flux $\mathfrak F^{(3)}$ and 
general NS-NS fluxes
can be written as 
\cite{Shelton:2005cf} (see also 
\cite{Berglund:2005dm,Aldazabal:2006up,Villadoro:2006ia,Shelton:2006fd,Micu:2007rd,Cassani:2007pq,Blumenhagen:2015kja}) 
\eq{
  \label{sp}
  W = \int_{\mathcal X} \Bigl( \mathfrak F^{(3)} + 
  \mathcal D \op \Phi^{\rm ev}_c \Bigr) \wedge \Omega_3\,,
}
with the following form of the twisted differential $\mathcal D$
\eq{
  \label{d_operator_01}
  \mathcal D = d - H\wedge\: - F\circ\: - Q\bullet\: - R\,\llcorner \,.
}
Expanding the forms in the bases introduced above and integrating, gives a  more accessible form of the superpotential
\eq{
\label{thebigW}
   W=&
  -\bigl({\mathfrak f}_\lambda  X^\lambda -\tilde {\mathfrak f}^\lambda  F_\lambda  \bigr) 
  +i\op S \big( h_\lambda  X^\lambda - \tilde h^\lambda  F_\lambda \bigr) \\[4pt]
  &+i\op G^a  \bigl( f_\lambda{}_a   X^\lambda - \tilde f^\lambda{}_a  F_\lambda \bigr) 
  -i\op T_{\alpha}  \bigl( q_\lambda{}^\alpha   X^\lambda - \tilde q^\lambda{}^\alpha  F_\lambda\bigr)\,.
}
The parameters ${\mathfrak f}_\lambda$, $\tilde{\mathfrak f}^\lambda$, $\tilde{h}^\lambda$, 
$f_\lambda{}_a$, $\tilde f^\lambda{}_a$,  $q_\lambda{}^\alpha$ and $\tilde q^\lambda{}^\alpha$ are the components of the fluxes in \eqref{sp} and \eqref{d_operator_01}, see \cite{Blumenhagen:2015kja} for precise definitions.
These fluxes are subject to Bianchi identities arising from the nilpotency condition ${\mathcal D}^2=0$. The $R$-flux does not
appear in $W$ as it is eliminated by the orientifold projection. The non-geometric fluxes $q_\lambda{}^\alpha$ and $\tilde q^\lambda{}^\alpha$ enter only in the last line and make it possible to stabilize the K\"ahler moduli $T_\alpha$ at tree-level. 

With the superpotential and the K\"ahler potential at hand we can compute the F-term
scalar potential 
\eq{
  \label{VF}
  V_F = \frac{M_\text{Pl}^4}{4 \pi}\,\, e^{K} \Bigl( K^{I\ov J} D_IW D_{\ov J}\ov W - 3 \op\bigl|W\bigr|^2 \Bigr) \, ,
}
where  $K_{I\ov J} = \partial_I \partial_{\ov J}\op K$ and
$D_I W = \partial_I W + (\partial_I K)\op W$, and the indices run over the moduli fields. As  is well known, the supersymmetric vacua of this scalar potential are determined  by 
 $F_I=D_I W = 0$ which implies $\partial_I V = 0$. 
It was shown in \cite{Blumenhagen:2015lta} that the orientifold projected potential of ${\cal N}=2$ GSUGRA splits 
into three terms 
\eq{
\label{potentialfull}
         V=V_F+V_D+V^{\rm NS}_{\rm tad} \,,
}
where $V_F$ is precisely the F-term scalar potential \eqref{VF}. $V^{\rm NS}_{\rm tad}$
is the NS-NS tadpole that will be canceled against the tension of 
branes and orientifold planes, once R-R tadpole cancellation is taken
into account. $V_D$ is an additional D-term potential \cite{Robbins:2007yv}
that results from the abelian gauge fields for $h^{2,1}_+>0$ and takes the form
\eq{
  \label{dtermpot}
   V_D &= -\frac{M^4_{\rm Pl}}{2} \, \Bigl[ ({\rm Im}\, \mathcal N)^{-1} \Bigr]^{\hat\lambda\hat\sigma} \op
   D_{\hat\lambda} \op D_{\hat\sigma} \,,
}
with
\eq{
\label{res_054}
D_{\hat{\lambda}} =
 \frac{1}{{\mathcal V}}\left[ 
  -r_{\hat\lambda}\, \big( e^\phi {\cal V}-\tfrac{1}{2}\op \kappa_{ \alpha a b} \op
    t^\alpha b^a b^b\Big) - q_{\hat\lambda}{}^{a}\, \kappa_{a \alpha  b}\op
     t^\alpha b^b + f_{{\hat\lambda} \alpha}\, t^\alpha \right]\,.
}
Here  ${\mathcal N}_{\hat \lambda \hat \sigma}$ are the period matrix elements with $\hat\sigma,\, \hat\lambda=1, \ldots, h^{2,1}_+$.
The potential contains $R$-flux $r_{\hat\lambda}$, $Q$-flux $q_{\hat\lambda}{}^{a}$ and geometric flux $f_{{\hat\lambda} \alpha}$. 

Finally, let us provide the expressions for the various mass scales.
The string and KK scale are
\eq{
\label{stringandKKscale}
       M_{\rm s}= {\sqrt{\pi} M_{\rm Pl}\over s^{1\over 4}\,{\cal V}^{1\over 2}}\,, 
       \hspace{40pt}
       M_{\rm KK}= {M_{\rm Pl}\over \sqrt{4\pi}\, {\cal V}^{2\over 3}} \, .
}
It is also useful to introduce the gravitino mass, which can be used to estimate the supersymmetry breaking scale.
This mass is given by
\eq{
\label{m32}
M^2_{\frac32} = e^{K}\op  |W|^2 \op {M_{\rm Pl}^2\over {4\pi}}\, .
}
In these expressions it is understood that the moduli VEVs, as well as $K$ and $W$, are evaluated at the minimum.

\section{The flux scaling scenario and its phenomenology}

Utilizing the structure reviewed in the last section, 
we now present  the flux-scaling scenario for moduli stabilization introduced in 
\cite{Blumenhagen:2015kja}. One important result of this work is that by turning on $n+1$ 
fluxes for $n$ moduli,  an F-term scalar potential that admits non-supersymmetric
AdS minima of scaling type is induced. Scaling means that in these minima the moduli 
vacuum expectation values (VEVs) and masses
are determined by ratios of products of  fluxes, thereby leading to parametric control of these quantities.
This is crucial to ensure that the moduli
are stabilized  in their perturbative regime while the moduli masses are separated from the string and Kaluza-Klein
scales. In this way the self-consistency of the moduli stabilization scheme can be justified.

To illustrate the main features of the flux-scaling minima we present in some detail a simple prototype.
We then discuss some general physical and phenomenological properties of these types of models.

\subsection{Flux Scaling Vacua}
\label{toymod}
We start by considering a representative example in which the underlying
geometry  is an isotropic torus with fixed complex structure, and where the orientifold 
projection is chosen such that  $h^{1,1}_- = 0$.
This implies that the $U$ and $G^a$ moduli are absent. The remaining moduli are the axio-dilaton $S = s + i\op c$ and one K\"ahler modulus $T = \tau + i \rho$. The imaginary parts $c$ and $\rho$ are axions, while $s$ and $\tau$ denote the saxions. 
Note that $s = e^{-\phi}$ gives the string coupling whereas $\tau = {\cal V}^{2 \over 3}$ encodes the volume of the isotropic torus. 
The K\"ahler potential for this geometry is
\eq{K=-\log (S+ \ov S)-3\, \log (T + \ov T)\,,}
and for the superpotential we choose
\eq{ \label{SPtoymodel}
W=i\,{\mathfrak f}+i\,h\,S+i\,q\,T \, .
} 
Since all moduli appear in the superpotential, we can stabilize them at tree-level without invoking subleading instanton-, gaugino- 
or $\alpha'$-corrections,  as it is needed for instance in the KKLT \cite{Kachru:2003aw} or the large volume scenario \cite{Balasubramanian:2005zx}. 
Next, notice 
that $W$ only depends on the axionic combination $\theta = hc + q \rho$ and not on the orthogonal one.
To gain further intuition, let us first investigate the supersymmetric vacuum with $D_I W = 0$. Schematically the moduli are fixed as
\eq{
\theta = hc + q \rho = 0, \qquad i h s \sim W, \qquad iq\tau \sim W \, .
}
Plugging this back into \eqref{SPtoymodel} then determines each term to scale individually as $W\sim i{\mathfrak f}$, so that
$s \sim {{\mathfrak f} \over h}$ and $\tau \sim   {{\mathfrak f} \over q}$. This scaling holds in all minima of this model and is the defining property of the class of flux-scaling vacua.
At the minimum there is an overall flux dependence dictated by the superpotential $W$. This then fixes all interesting physical parameters.
Concretely:
\begin{itemize}
\item the flux dependence of the moduli, here $s \sim {{\mathfrak f} \over h}$ and $\tau \sim   {{\mathfrak f} \over q}$, 
\item the (negative) cosmological constant $V \sim e^K |W|^2 \sim - {h q^3 \over {\mathfrak f}^2}$, 
\item the moduli masses $M^2_{{\rm mod}} \sim (\partial \partial K)^{-1} \partial \partial V \sim V $,
\item the supersymmetry breaking scale $M_{3 \over 2} \sim e^{K\over 2} |W|  \sim M_{{\rm mod}}$,
\item the string scale $M_{\rm s} \sim s^{-{1 \over 4}} {\cal V}^{-{1 \over 2}} \sim {h^{1 \over 4} q^{3 \over 4} \over {\mathfrak f}}$,
\item and the Kaluza-Klein scale $M_{{\rm KK}} \sim {\cal V}^{- {2 \over 3}} \sim {q \over {\mathfrak f}}$.
\end{itemize}
In table \ref{table_extr} we display all minima of the model, where the sign
of the R-R flux ${\mathfrak f}$ has to be chosen such that the saxions
come out positive.
\begin{table}[t]
      \label{table_extr}
  \centering
  \renewcommand{\arraystretch}{1.3}
  \begin{tabular}{|c|c|c|c|c|}
  \hline
   solution& $(s,\tau,\theta)$ & susy &  tachyons & $\Lambda$ \\
  \hline\hline
  & & & &\\[-0.5cm]
  1 & $(-{{\mathfrak f}\over 2 h}, -{3 {\mathfrak f}\over 2 q},0)$ & yes & yes & AdS \\[0.2cm]
  2 & $({{\mathfrak f}\over 8 h}, {3 {\mathfrak f}\over 8 q},0)$ & no & yes & AdS \\[0.2cm]
  3 & $(-{{\mathfrak f}\over h}, -{6 {\mathfrak f}\over 5 q},0)$ & no & no & AdS \\[0.2cm]
 \hline
     \end{tabular} 
     \caption{\small Extrema of the scalar potential in the prototype example.}
\end{table}
There is indeed a fully stable non-supersymmetric AdS vacuum for which the masses are  
\eq{ \label{massessimplemodel}
M_{{\rm mod},i}^2=\mu_i  \op {h q^3\over \tilde{\mathfrak f}^2}\, {M_{\rm Pl}^2 \over 4 \pi \cdot 2^4} \,,
}
with the numerical values 
\eq{ 
\label{massesA}
\mu_i \, \approx \, (6.2, 1.7\, ; \, 3.4, 0)\,.
}
The first (last) two masses correspond to combinations of saxions (axions).
Let us now summarize the physical aspects of these flux-scaling minima.
\begin{itemize}

\item We gain control over string loop and $\alpha'$-corrections by an appropriate choice of the vacuum expectation values  of the saxions. 
Flux-scaling minima allow to choose them in the large volume, large complex structure and weak string coupling regime. 
In our example the VEVs are given by $e^{-\phi} = s \sim  \frac{f}{h}$ and ${\cal V}^{2 \over 3} = \tau \sim  \frac{f}{q}$. Hence, for $f>h,\,q$, we are in the desired region of control.

\item Due to massless axions, the supersymmetric AdS vacua of flux-scaling type contain tachyons as predicted by the aforementioned no-go theorem \cite{Conlon:2006tq}. 
Nevertheless, also stable AdS vacua are present with supersymmetry broken at a high scale $M_{3 \over 2} \sim M_{{\rm mod}}$.
Choosing the saxions in the physical regime, the vacua are always AdS.

\item All masses have the same flux dependence. The original goal to get parametric control over an axion mass to make it lighter for axion monodromy inflation is therefore not yet reached, although in many models an axion was numerically lighter than the other moduli.

\item As explained in the introduction, the hierarchy 
\eq{ 
\label{masshierb}
M_{\rm Pl}> M_{\rm s}>M_{\rm KK}> M_{\rm mod}
}
among the Planck-scale, the string-scale, the Kaluza-Klein-scale and the moduli masses must hold as otherwise the massive KK and string modes cannot be integrated out.
Recall that all scales should differ by only a factor of $\mathcal O(10)$, which makes this a difficult task. In our prototype example the KK scale comes out to be of the same
order as the moduli masses. However, in \cite{Blumenhagen:2015kja} models were designed
where the hierarchy \eqref{masshierb} could be realized.

\item The gravitino mass  $M_{3\over2}$ is always of the same order as the moduli masses $M_{3 \over 2} \sim M_{{\rm mod}}$. For moderate values of the fluxes,
we therefore have a high supersymmetry
 breaking scale at around $10^{14}\text{-}10^{15}$GeV. To lower the supersymmetry
breaking scale in the visible Standard-model sector, it has to be sequestered
from the bulk. Realizing  the SM on a stack of D7-branes wrapping 
a four-cycle,  the auxiliary field residing in the same superfield must
satisfy $F^{\rm SM}=0$. Then the expectation is that the dominant source
of susy breaking is anomaly mediation leading generically to a further suppression
by a loop-factor so that one gets  soft terms in the intermediate regime of $10^{11}\,$GeV.

\item Usually in the given framework the fluxes are considered  as a perturbation around a Calabi-Yau background. To make this self-consistent, the backreaction of the fluxes onto the geometry must be under control. In case of only NS-NS and R-R three-form fluxes the backreaction leads to a warp factor \cite{Giddings:2001yu}. 
The backreaction is diluted in the large volume limit which can be taken due to the non-stabilization of the K\"ahler moduli.
We showed that non-geometric fluxes have a backreaction of ${\mathcal O}(1)$ while not being substantially large. Although not ruled out, the minima are therefore on less firm ground than purely geometric vacua.

\item In most vacua, the tadpole contribution of the fluxes is negative. This is in agreement with an earlier analysis in the T-dual type IIA setup \cite{Camara:2005dc,Aldazabal:2007sn,Aldazabal:2006up}. The tadpoles can then be canceled by 
D3- or D7-branes instead of O3- and O7-planes.

\end{itemize}

\subsection{Tachyon Uplift}

In models with a larger number of moduli, extrema often have tachyonic directions. For the K\"ahler moduli sector we were able to find a natural mechanism to lift theses directions using  a stack of $N$ D7-branes wrapping an internal four-cycle $\Sigma$. These branes carry a $U(N)$ gauge flux whose $U(1) \subset U(N)$ part 
contributes a Fayet-Iliopoulos (FI) D-term to the scalar potential
\eq{
V_D = {M^4_{Pl} \over 2 Re(f)}\xi^2 \, , 
\hspace{50pt}
 \xi = {1 \over {\cal V}} \int_{\Sigma} J \wedge c_1(L).
}
Here ${\cal V}$ is the internal Volume, $J$ is the K\"ahler form and $c_1(L)$ is the first Chern-class of the $U(1)$ line bundle, namely its field strength tensor $F$. 
For D7-branes the gauge kinetic function  $f$ is given by $f = T + \chi S$ with the instanton number $\chi = {1 \over 4\pi} \int F \wedge F$. 

From the St\"uckelberg mass term for the gauge field on the D7-brane
one can derive the gauging of the axionic shift symmetries. Since the superpotential does contain the axions, 
its gauge invariance leads to conditions on the allowed closed string  fluxes,
if magnetized  D7-branes are present, as well. They can be considered
as generalized Freed-Witten anomaly conditions.
These conditions imply a simplification of  the above FI-term so that 
it vanishes  not only  in the supersymmetric minimum, as expected, but also in the nearly stable non-supersymmetric minimum. Hence, it does not shift the existing minima but
its only effect on the old vacuum is a positive contribution to the mass of the K\"ahler tachyons. The square of these masses can become positive while the other moduli masses are unaffected. 
We note that this mass uplift mechanism does not work for complex structure moduli,
simply  because they do not enter in the D7-brane D-term.

\subsection{De Sitter Uplift}

Stable AdS vacua need to be uplifted to de Sitter to provide a realistic framework for inflation. 
In the following, we consider two methods to add a positive contribution to the potential. The first one is by adding an $\ov {\rm D3}$-brane while the second approach is by adding the D-term potential \eqref{dtermpot}.
As shown explicitly in \cite{Blumenhagen:2015xpa}, both methods are successful in generating Minkowski or de Sitter flux-scaling vacua. 
We discuss them in turn.
\begin{itemize}

\item \textbf{$\ov{{\rm D}3}$-branes}\\
Following the usual uplift procedure, 
one could take an  existing AdS minimum and add an $\overline{{\rm D}3}$-brane at a warped throat to uplift the initial flux-scaling minimum to de Sitter. 
For our models, this procedure turned out to fail. The vacua are destabilized before
the cosmological constant changes sign.
Hence a continuous  uplift does not work. Instead one can look for \textit{new} flux-scaling minima of the full potential where the 
$\ov {{\rm D}3}$ brane is included a priori. To this end, we add to the F-term potential the
positive semi-definite uplift term 
\eq{
\label{vupgen}
V_{\rm up} = \frac{A}{ \mathcal{V}^{4\over 3}}\, {M^4_{\rm Pl}\over 4\pi}\, , 
}
with $A$ a positive constant depending on the warp factor in the throat. Then we 
try to solve the conditions $\partial_I V=0$ and $V=0$ in terms of the
moduli VEVs and the parameter $A$. If a Minkowski minimum exists,
then choosing a slightly larger value for the parameter $A$ leads
to a dS vacuum.

\item \textbf{Non-geometric D-term}\\
As we discussed, for orientifold projections with non-zero $h^{2,1}_+$, 
the dimensional reduction of the R-R $C^{(4)}$-form leads to 
abelian gauge fields in four-dimensions.  These do feature a  
positive semi-definite D-term potential $V_D$ that only depends on the
saxions. Taking the constraints from the Bianchi identities into account, 
one can now also search directly  for new flux-scaling Minkowski/dS minima.  

\end{itemize}


\section{Inflation}

As explained in the introduction, our initial motivation for investigating the non-geometric superpotential was to realize high-scale  single-field F-term axion monodromy inflation. In this type of inflation a single axion becomes the inflaton, while the other moduli do not participate in inflation and should therefore be heavier than the Hubble scale. Our goal is to gain control over this hierarchy in a concrete framework of string moduli stabilization.
First,  we present a simple, analytically  solvable model featuring  parametric control over an axion mass via a two-step procedure. Second, for this set-up we study inflation.
For different choices of a flux parameter, this leads  to linear, quadratic or Starobinski-like inflation. 

\subsection{Realizing a single light axion}
\label{twostep}

Let us try to obtain a parametrically controlled -- but not accidental -- hierarchy between one axion and the other moduli. We will follow the ansatz of \cite{Blumenhagen:2014nba} which works in two steps. The first step is to turn on a superpotential $\lambda W_0$
to stabilize all moduli except an axion which we identify as the inflaton. The vacuum of our choice is then perturbed by a small correction $f_{ax} \Delta W$ which stabilizes the remaining axion. The full superpotential is given by
\eq{
W=\lambda\, W_0+f_{ax}\,\Delta W\,.
}
A hierarchy between the inflaton mass $M_\theta$ and the remaining moduli masses is now obtained by choosing $\lambda \gg f_{ax}$.  As long as one restricts the  flux 
to obey Dirac's quantization condition, they are integers
and one can only get $\lambda \gg |f_{ax}|$ by choosing $\lambda$ large. 
Since the moduli masses scale like $M_{\rm mod}\sim\lambda$ while
the KK and the string  scale are independent of $\lambda$, 
it is hard (presumably even impossible)  to keep the hierarchy $M_{\rm KK}> M_{\rm mod}$.
Recall that there was only a difference of ${\mathcal O}(10)$ between the moduli masses and the KK scale. Scaling the moduli masses up to separate them from the inflaton mass necessarily leads to a conflict. 

As observed in \cite{Blumenhagen:2014nba}, the  polynomial corrections to the complex
structure prepotential \eqref{f_corr} can be absorbed in a redefinition
of the fluxes. 
Indeed,  when evaluating  the superpotential \eqref{thebigW}, 
the  corrections $a_{ij}$ and $b_i$ can  be incorporated into 
shifts in the fluxes. For instance, for the $H$-fluxes this means
\eq{
\label{fluxshift}
     h_0 = h_0 -b_i\op \tilde h^i  \,,  \qquad h_i = h_i -a_{ij}
     \op\tilde h^j - b_i \op\tilde h^0 \,,
}
with $i\in\{1,\dots,h^{2,1}_-\}$ and analogously for the other fluxes.
Recall that the purely imaginary contribution $i\op\gamma$ corresponds
to  $\alpha'$-corrections to the  K\"ahler potential for the 
K\"ahler moduli in a mirror-dual setting. In the large
complex-structure regime we are 
employing here, these corrections can be neglected. 
Similarly, in this regime also the 
non-perturbative corrections $F_{\rm inst.}$ are negligible. 
Since on a genuine CY  the coefficients, $a_{ij}$ and $b_i$, are fixed rational
numbers, the effective fluxes are not necessarily integers. 
It was shown in \cite{Blumenhagen:2015xpa}, that this relaxation provides more freedom
for finding models realizing the hierarchy \eqref{introhierarchy} already for moderate values of $\lambda$. These models need fluxes with values smaller than one.

\subsection{A toy model of inflation}
\label{arielle}

As described in the previous section \ref{twostep}, there exists a way to construct
models in which an axion is the lightest mode,  though at the expense
of allowing rational values of the fluxes smaller than one. 
Following \cite{Blumenhagen:2015qda}, let us now  consider a simple
toy model of  inflation that, as far as we can see, captures already many
of the qualitative features that will show up in more elaborate models.

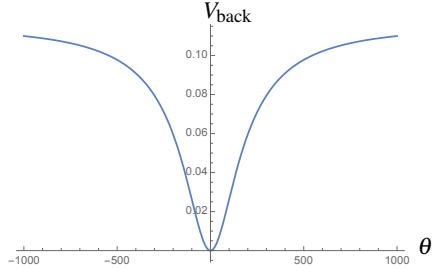
\begin{figure}[t]
  \centering
  \vspace{0.4cm}
  \includegraphics[width=0.35\textwidth]{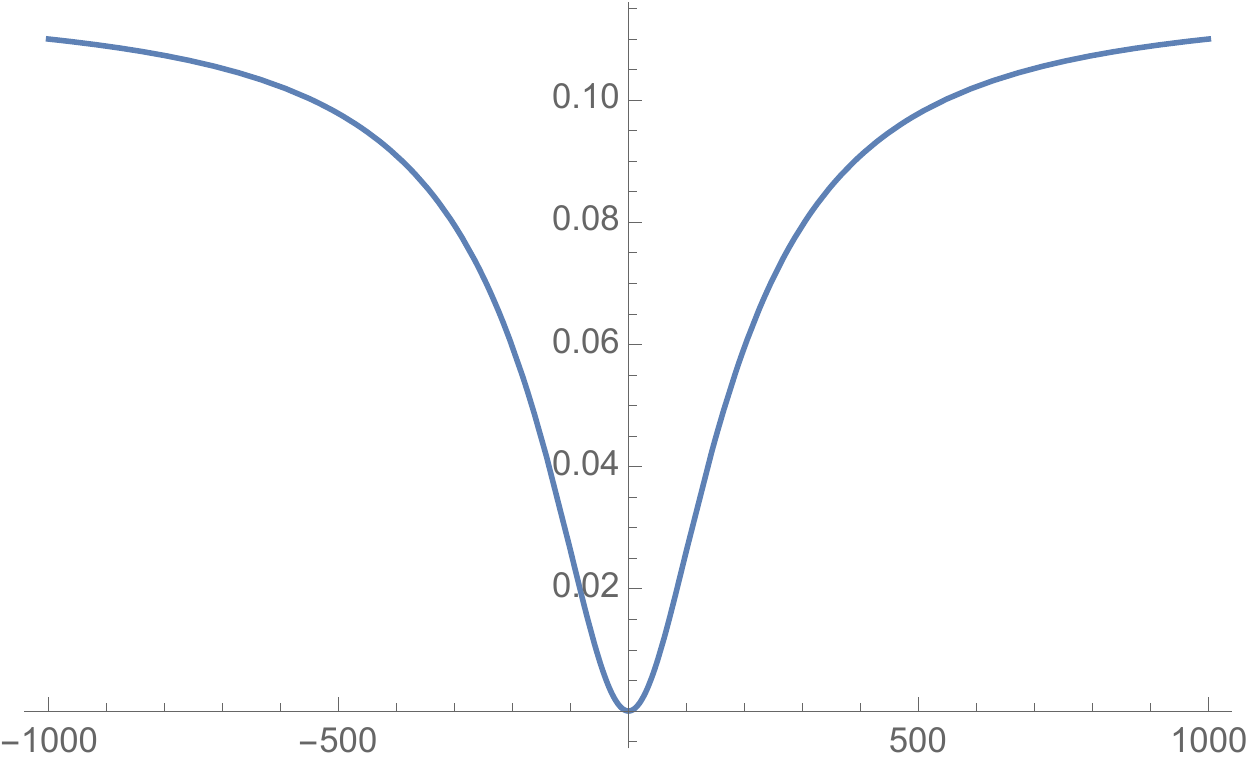}
  \begin{picture}(0,0)
  \put(0,6){\footnotesize$\theta$}
  \put(-80,95){\footnotesize $V_{\rm back}$}
  \end{picture}
  \caption{
  \small The potential $V_{\rm back}(\theta)$ in units of
    $M^4_{\rm Pl}/(4\pi)$ for fluxes $h=1$, $q=1$, $\tilde{\mathfrak
    f}=10$ and $\lambda=10$. One can see the different regions of inflation, quadratic, linear and plateau inflation.
      }
  \label{bild1}
\end{figure}

For this purpose we just take the simple toy model of section \ref{toymod} with the saxionic moduli $s$, $\tau$ and the axion $\theta = hc + q \rho$, which will serve as the inflaton. 
By placing (by hand) a parameter $\lambda$ into the potential
\eq{
 V \sim \,\lambda^2 \, \left[
{(hs - \tilde{\mathfrak f} )^2 \over s \tau^3} - {6 h q s+ 2 q \tilde{\mathfrak f}\over
  s \tau^2} - {5 q^2\over 3 s \tau} \right]  +{\theta^2  \over s \tau^3} \,,
  }
we obtain a controllable hierarchy between the inflaton $\theta$ and the other moduli $s$ and $\tau$, manifest in $M_{s, \tau} / M_\theta \sim \lambda$. We now consider a slowly rolling inflaton with the other moduli $s$, $\tau$ adiabatically adjusting. Solving the conditions $\partial_s V = \partial_{\tau} V= 0$ for non-vanishing $\theta$ gives
\eq{
\label{backshift}
    \tau_0(\theta)={3\over 20\op q} \left(  4 \op\tilde{\mathfrak f}
      + \sqrt{10\op \left( \tfrac{\theta}{\lambda}\right)^2 +16\op
        \tilde{\mathfrak f}^2} \:\right), \qquad 
    s_0(\theta)={1\over 4\op h}  \,\sqrt{10\op \left( \tfrac{\theta}{\lambda}\right)^2 +16\op  \tilde{\mathfrak f}^2} \, .
}
Plugging these values back into the potential and assuming for simplicity a constant
uplifting to Minkowski gives the so-called backreacted effective inflation potential
for $\theta$ \cite{Dong:2010in}  
\eq{
\label{backpotential}
  V_{\rm back}(\theta)=
  \frac{25 \lambda^2 h q^3}{108\op \tilde{\mathfrak f}^2} \op
  \frac{ 5 \bigl( \frac{\theta}{\lambda}\bigr)^2 - 4 \op\tilde{\mathfrak f}\op \Big(  4\op \tilde{\mathfrak f} - 
  \sqrt{
  10\bigl( \frac{\theta}{\lambda}\bigr)^2 +16\op \tilde{\mathfrak f}^2}  \Bigr)
  }{ 
  \Big(  4\op \tilde{\mathfrak f} + \sqrt{
    10\op\left( {\theta\over \lambda}\right)^2 +16\op \tilde{\mathfrak f}^2
  }  \Big)^2} ,
} 
which is depicted in figure \ref{bild1}. 
We notice that the inflaton potential changes from a convex behavior
near zero to a concave one for $\theta/\lambda\gg \tilde{\mathfrak f}$.
As a consequence, for a large hierarchy with $\lambda = 60$ it turns out that enough e-foldings are collected in the quadratic looking region close to $\theta=0$. The trajectory \eqref{backshift} is a good approximation as only the inflaton is below the Hubble scale. The quadratic term dominates giving a tensor-to-scalar ratio of around $r \sim 0.133$.

When looking at a smaller hierarchy with around $\lambda = 10$ the backreaction becomes substantial and most of the e-foldings are collected at the linear part of the potential. The tensor-to-scalar ratio is therefore expected to drop down to $r \sim 0.08$ as in linear inflation. Still the other moduli are above the Hubble scale but the mass gap is so low that one might get corrections from a full multifield analysis that could also correct the trajectory \eqref{backshift}.

Finally we consider  the case $\lambda = 1$ when there is no mass hierarchy at all. Now all moduli have to be taken into account, as all their masses are around or lower than the Hubble scale. By solving the equations of motion numerically for the field trajectories, we see that indeed all moduli $s$, $\tau$ and $\theta$ take part in inflation. Inflation occurs at a plateau resulting in a tensor-to-scalar ratio of around $r \sim 0.0015$, a typical value for a plateau potential. A more detailed analysis would necessarily require a multi-field treatment.


\section{Conclusions}

In this proceedings article we have reviewed some recent developments in the field of string moduli
stabilization and its application to string cosmology. This work
was motivated by,  but is not necessarily restricted to, providing a treatable
framework for realizing single field F-term axion monodromy inflation in
string theory. 

The conceptual framework is that of
DFT compactified on orientifolded Calabi-Yau three-folds equipped with additional geometric
and non-geometric fluxes, leading to precisely the scalar potential
of orientifolded ${\cal N}=2$ GSUGRA. We mentioned that, since the backreaction
of these fluxes is of order $\mathcal O(1)$, we cannot argue via  a dilute flux limit that
the vacua found in this four-dimensional approach do truly lift up
to solutions of the ten-dimensional string or DFT equations of motion.

The salient new feature of our approach is that all closed string moduli
are stabilized at tree-level by non-geometric fluxes. Moreover, in this scheme of 
moduli stabilization we have  managed
to identify a special class of non-supersymme\-tric scaling type minima,
that allow parametric control over the relevant mass scales.
Many questions relevant for string phenomenology, such as soft
supersymmetry breaking terms or a possible uplift to Minkowski/dS vacua,
were  discussed, though the main motivation comes from the application to inflationary
cosmology.

For a toy model of axion inflation, we showed
that the inflationary trajectory interpolates between quadratic, linear and plateau-like
inflation.

\vspace{0.7cm}
\emph{Acknowledgments:}
We are grateful to Cesar Damian, Yuta Sekiguchi, Rui Sun and Florian Wolf for collaboration on topics reviewed in this paper. A.F. thanks the Max-Planck-Institut f\"ur Physik and the Ludwig-Maximillians-Universit\"at for hospitality and support. 


\end{document}